# Approximate Backbone Based Multilevel Algorithm for Next Release Problem

He Jiang
School of Software
Dalian University of Technology
Dalian 116621, China
jianghe@dlut.edu.cn

Jifeng Xuan
School of Mathematical Sciences
Dalian University of Technology
Dalian 116024, China
xuan@mail.dlut.edu.cn

Zhilei Ren
School of Mathematical Sciences
Dalian University of Technology
Dalian 116024, China
ren@mail.dlut.edu.cn

## ABSTRACT

The next release problem (NRP) aims to effectively select software requirements in order to acquire maximum customer profits. As an NP-hard problem in software requirement engineering, NRP lacks efficient approximate algorithms for large scale instances. The backbone is a new tool for tackling large scale NP-hard problems in recent years. In this paper, we employ the backbone to design high performance approximate algorithms for large scale NRP instances. Firstly we show that it is NP-hard to obtain the backbone of NRP. Then, we illustrate by fitness landscape analysis that the backbone can be well approximated by the shared common parts of local optimal solutions. Therefore, we propose an approximate backbone based multilevel algorithm (ABMA) to solve large scale NRP instances. This algorithm iteratively explores the search spaces by multilevel reductions and refinements. Experimental results demonstrate that ABMA outperforms existing algorithms on large instances in terms of solution quality and running time.

## Categories and Subject Descriptors

D.2.1 [**Software Engineering**]: Requirements/Specifications – *Methodologies*; I.2.8 [**Artificial Intelligence**]: Problem Solving, Control Methods, and Search - *Heuristic methods*

## 1. INTRODUCTION

The next release problem (NRP) is a combinatorial optimization problem in software requirement engineering proposed by Bagnall, et al. in 2001 [1]. This problem seeks to maximize the customer profits from a set of dependent requirements, under the constraint of a predefined budget bound. NRP and its variants have attracted much attention in requirement engineering, such as component selection and prioritization [2], multi-objective next release problem (MONRP) [5][11][22], and release planning [18][19]. There are numerous applications of NRP in requirement analysis [9][10][18]. Some further researches on NRP have been conducted, including fairness analysis in requirement assignment [5], multi-objective search based approaches for software project planning [8], and sensitivity analysis in requirement engineering [11].

NRP has been proved as "*NP−hard* even when it is basic and customer requirements are independent" [1], i.e., no exact algorithm exists to achieve global optimal solutions in polynomial time unless $P = NP$ [6]. In practice, it is straightforward to find approximate algorithms to obtain near-optimal solutions within polynomial time. In the literature, many approximate algorithms have been proposed for NRP and its variants, including greedy algorithms [1][11], greedy randomized adaptive search procedure (GRASP) [1], local searches (e.g., sampling hill climbing and simulated annealing) [1][2], genetic algorithm [19], etc. Among these algorithms, a simulated annealing algorithm by Lundy & Mees (LMSA) [14] is the best one for solving NRP approximately. LMSA can work efficiently on small instances of this problem, but there is an absence of approximate algorithms for large instances (an instance is generated by specifying particular values for all the parameters of a problem [6]).

As an effective tool for approximate algorithm design, the backbone has been one of the hot topics of research on *NP−hard* problems in recent years. The backbone is defined as the common parts of all global optimal solutions for a problem instance [3]. If the backbone is obtained, the global optimal solutions can be partly constructed. By fixing this part of global optimal solutions, the scale of the original instance can be reduced. Thus, an ideal approach is to obtain the backbone and then reduce the difficulty for solving a problem instance. However, since it is usually intractable to obtain the backbone of *NP−hard* problems, most of algorithms choose to construct the approximate backbone with the common part of local optimal solutions instead. Dubois & Dequen [3] investigate an approximate backbone based heuristic algorithm for solving the hard 3-satisfiability problem (3-SAT). Zhang, et al. [21] design a backbone based approximate algorithm for maximum satisfiability problem (Max-SAT). Kilby, et al. [13] develop an approximate backbone based algorithm for traveling salesman problem (TSP). Jiang, et al. [12] present an approximate backbone based ant colony algorithm for quadratic assignment problem (QAP). Apart from the algorithms for above classic optimization problems, there is only one backbone based algorithm for the problems in software engineering to our knowledge. Mahdavi, et al. [15] propose a "building block" based multiple hill climbing approach for software module clustering problem. In their approach, the concept of building blocks is similar to that of the approximate backbone when applying search based approaches to the problems in software engineering.

As a new problem in the family of $NP-hard$ problems, NRP has not been well investigated on either theoretical analysis or algorithm design. Since NRP is a practical problem in requirement engineering, it is necessary to develop an algorithm for large instances arising in real-world applications. In this paper, we propose an approximate backbone based multilevel algorithm (ABMA) for solving large NRP instances. In contrast to existing algorithms, our new algorithm can reduce and refine the instances by fixing the approximate backbone iteratively. Firstly, we prove that it is $NP-hard$ to obtain the backbone of NRP. In the proof, we map any instance to a biased instance with a unique optimal solution, which is also optimal to the original instance. Secondly, we present the similarity between local optimal solutions and global optimal solutions by fitness landscape analysis. Then the approximate backbone can be constructed with the common parts of local optimal solutions based on the similarity. Thirdly, ABMA is proposed to solve NRP. This algorithm includes two iterative phases: reduction and refinement. The reduction phase iteratively reduces the instance to obtain a new smaller instance by fixing the approximate backbone; the refinement phase combines the solution of this instance and the approximate backbones into a solution for the original instance. Since LMSA, the best approximate algorithm for NRP up till now, cannot work efficiently for large instances, we design greedy climbing search (GCS), a hill climbing operator based on a greedy strategy. GCS is incorporated into ABMA as a local search operator. Finally, experiments are conducted on extensive instances to evaluate the performance of our new algorithm. Experimental results show that ABMA can achieve better performance on large instances than existing algorithms.

The primary contributions of this paper are as follows:

First, this paper presents the theoretical analysis of the computational complexity for obtaining the backbone in NRP, i.e., it is $NP-hard$ to obtain the backbone of NRP.

Second, this paper shows a multilevel algorithm to reduce the scale of instances by fixing the approximate backbone. This algorithm can work well on large scale instances. Some similar strategies can also be applied to other $NP-hard$ problems.

Finally, this paper presents how to incorporate the backbone into an approximate algorithm for solving NRP. It is the first application of the backbone to requirement engineering. Some of the complex problems in requirement engineering can be approximately solved by the backbone based algorithms, especially $NP-hard$ problems.

The remainder of this paper is organized as follows. Section 2 states the related definitions of NRP. Section 3 presents the computational complexity results for the backbone. Section 4 introduces the approximate backbone based multilevel algorithms for NRP. Section 5 presents the experimental results. Section 6 briefly concludes this paper and points out the potential ways in future work.

## 2. PRELIMINARIES

In this section, we introduce the application scenario for NRP and then give some related definitions and properties.

When a software company decides to upgrade its software, many candidate requirements can be included in the next release (e.g., the version upgrading of a web browser, Google Chrome [7] or the version upgrading of an integrated development environment, Eclipse [4]). On one hand, it is usually too expensive to implement all the requirements for this software company. On the other hand, every customer may request a fraction of those candidate requirements and provides a potential commercial profit for the software company. When all the requirements requested by a customer have been implemented, the software company can gain the profit from this costumer. In addition, there may be some dependency relationships among those candidate requirements in a real-world software project, i.e., a requirement can only be implemented after some other ones. NRP aims to determine a subset of those candidate requirements under a predefined budget bound so that this company could achieve maximum profits from its customers.

According to this application scenario, we give the formal definitions of NRP as follows. In a software system, let $R$ be the set of all candidate requirements and the cardinality of $R$ is defined as $|R|=m$. Every requirement $r_i \in R$ ( $1 \le i \le m$ ) is associated with a nonnegative cost $c_i$. A directed acyclic graph $G=(R,E)$ denotes the dependency relationships among those requirements, where $R$ is the set of vertexes and $E$ is the set of arcs. An arc $(r',r) \in E$ indicates that requirement $r$ depends on $r'$. Let $parents(r)$ be the set of requirements, which can reach $r$ via one or more arcs. Obviously, all the requirements in $parents(r)$ must be listed in the development plan before $r$ is available in the next release.

Let $S=\{1,2,\cdots,n\}$ be all the customers related to those requirements. Every customer $i$ requests a set of requirements $R_i \subseteq R$. Let $w_i \in W$ be the profit gained from customer $i$. Let $parent(R_i)=\bigcup_{r \in R_i} partent(r)$ and $\hat{R}_i = R_i \cup parent(R_i)$. Under the above definitions, a customer $i$ can be satisfied, if and only if all the requirements in $\hat{R}_i$ are listed in the release. Let $cost(\hat{R}_i)=\sum_{r_j \in \hat{R}_i} c_j$ be the cost for satisfying customer $i$. Let $S' \subseteq S$ be a subset of customers satisfied. The cost of $S'$ is defined as $cost(S')=cost(\bigcup_{i \in S'} \hat{R}_i)$ and the overall profit obtained is $\omega(S')=\sum_{i \in S'} w_i$.

Given an NRP instance (denoted as $NRP(S,R,W)$ ), a feasible solution is a subset $S' \subseteq S$ subject to $cost(S') < B$ , where $B$ is a predefined development budget bound. To facilitate the following discussion, we also formulate a feasible solution as a set of ordered pairs. For a feasible solution $S' \subseteq S$ , its ordered pair form is defined as $X=\{(i,b)|i \in S',b=1 \text{ or } i \notin S',b=0\}$ . Similarly, we also define $cost(X)=cost(\bigcup_{(i,1)\in X} \hat{R}_i)$ and $\omega(X)=\sum_{(i,1)\in X} w_i$ . Obviously, it is easy to convert $X$ and $S'$ into each other. Let $F_B$ be the set of all the feasible solutions for an instance $NRP(S,R,W)$ . The goal of NRP is to find a feasible solution $X^* \in F_B$ such that $\omega(X^*) = \max_{X \in F_B} \omega(X)$ .

Given an NRP instance $NRP(S,R,W)$, let $\Pi = \{X_1^*, X_2^*, \cdots, X_t^*\}$ be the set of all global optimal solutions. The backbone of $NRP(S,R,W)$ is defined as $bone(S,R,W) = \bigcap_{i=1}^{t} X_i^*$. Given an NRP instance $NRP(S,R,W)$, its biased instance is defined as $NRP(S,R,\hat{W})$, where $\hat{W} = \{\hat{w}_i \mid \hat{w}_i = w_i + 1/2^i, i \in S\}$. In other words, the biased instance can be viewed as an NRP instance with noise profits. Obviously, it takes $O(n)$ running time to construct the biased instance for an NRP instance.

In the following part, a simple NRP instance (this example is extracted from the data of a communication company [18]) is illustrated with 3 customers and 8 requirements. Table 1 shows the descriptions of these 8 requirements. Figure 1 shows the dependency relationships and the requirements requested by customers, where the arrows from above to below indicate the dependency relationships. For requirement set $R = \{r_1, r_2, ..., r_8\}$, let the cost of these requirements $c_1, c_2, ..., c_8$ be 6,10,16,4,1,7,6,1, respectively.

**Table 1. Requirements of a communication company**

| Requirement | Description | Cost |
|---|---|---|
| 1 | Cost Reduction of Transceiver | 6 |
| 2 | Expand Memory on BTS Controller | 10 |
| 3 | FCC Out-of-Band Emissions | 16 |
| 4 | Software Quality Initiative | 4 |
| 5 | USEast Inc. Feature 1 | 1 |
| 6 | USEast Inc. Feature 2 | 7 |
| 7 | China Feature 1 | 6 |
| 8 | China Feature 2 | 1 |

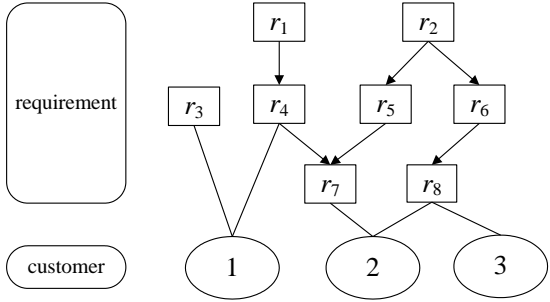

**Figure 1. Dependency of customers and requirements**

Given a cost ratio 0.7, the budget bound $B = 0.7 \bullet \sum_{r_i \in R} c_i \approx 36$ (simplified to be an integer). The requirements requested by the 3 customers are $R_1 = \{r_3, r_4\}, R_2 = \{r_7, r_8\}, R_3 = \{r_6\}$ and the profits of them $w_1, w_2, w_3$ are 30,25,20, respectively. Thus, the total requirements requested are $\hat{R}_1 = \{r_1, r_3, r_4\}$, $\hat{R}_2 = \{r_1, r_2, r_4, r_5, r_6, r_7, r_8\}$, $\hat{R}_3 = \{r_2, r_6, r_8\}$.

According to the definition of NRP, the profit and cost of the feasible solution $X_1 = \{(1,1),(2,0),(3,0)\}$ are 30 and 26, respectively. Similarly, the profit and cost of $X_2 = \{(1,0),(2,1),(3,1)\}$ are 45 and 35. Obviously, $X_2$ is better than $X_1$. However, $X_3 = \{(1,1),(2,1),(3,0)\}$ is unfeasible, because its cost 51 exceeds the bound $B$.

# 3. THEORETICAL ANALYSIS

According to the above definitions, we present the computational complexity analysis for achieving the backbone of NRP instance in this section. Due to the paper length limit, the proofs of Lemma 1 and Lemma 2 are not given in this paper.

**Lemma 1.** Given an NRP instance $NRP(S,R,W)$, if $w_i \in Z^+$ for any $i \in S$, then the biased NRP instance $NRP(S,R,\hat{W})$ has a unique global optimal solution.

**Lemma 2.** Given an NRP instance $NRP(S,R,W)$, if $w_i \in Z^+$ for any $i \in S$, then the unique global optimal solution of the biased NRP instance $NRP(S,R,\hat{W})$ is also a global optimal solution of $NRP(S,R,W)$.

**Theorem 1.** Unless $P = NP$, there exists no polynomial time algorithm to obtain the backbone of NRP.

**Proof.** (Proof by contradiction) We assume that this theorem is false. There must exist an algorithm $\pi$, which can obtain the backbone $bone(S,R,W)$ of NRP within polynomial time (denoted as $O(f(n))$, where $f(n)$ is a polynomial function of $n$.

Given any NRP instance $NRP(S,R,W)$, we assume that $w_i \in Z^+$ for any $i \in S$. This assumption is based on the above two lemmas. If there is an instance with $w_i \notin Z^+$, we can obtain a new instance with profits $w_i' \in Z^+$ by multiplying all the original profits by the same number. The solution of the new instance will be the same as that of the original one.

Now we can construct an algorithm to solve $NRP(S,R,W)$ as follows.

(a) We construct the biased instance $NRP(S,R,\hat{W})$ for $NRP(S,R,W)$ in $O(n)$ running time;

(b) According to the assumption, since $NRP(S,R,\hat{W})$ is also an instance of NRP, its backbone $bone(S,R,\hat{W})$ can be achieved within $O(f(n))$ running time by the algorithm $\pi$;

(c) By Lemma 1, $NRP(S,R,\hat{W})$ is an instance with unique global optimal solution. Thus, $bone(S,R,\hat{W})$ is a global optimal solution of $NRP(S,R,\hat{W})$;

(d) By Lemma 2, the global optimal solution of $NRP(S,R,\hat{W})$ is also a global optimal solution of $NRP(S,R,W)$. Thus, $bone(S,R,\hat{W})$ is also a global optimal solution of $NRP(S,R,W)$.

Therefore, the global optimal solution of $NRP(S,R,W)$ can be obtained within $O(n) + O(f(n))$ running time. This contradicts with the fact that NRP is $NP-hard$. Thus, this theorem is proved.

# 4. ABMA

According to the definition of the backbone, if the backbone of an NRP instance is obtained, the global optimal solutions can be partly constructed and the original instance can be reduced by fixing the backbone. However, as shown in Section 3, the backbone of NRP cannot be exactly obtained by a polynomial time algorithm. In this section, we firstly analyze the relationship between global and local optimal solutions by fitness landscape. Then, we show the ABMA algorithm for NRP. Finally, we present the greedy climbing search operator employed in ABMA.

## 4.1 Fitness Landscape Analysis

We conduct fitness landscape analysis [16] to investigate the relationship between global optimal solutions and local optimal solutions. For an instance, a global optimal solution is the best solution in the whole solution space and a local optimal solution is the best one in a specified neighborhood [17]. Usually, a local optimal solution can be returned by a local search algorithm within polynomial time. In addition, a local search algorithm can be called as a local search operator when it is incorporated into another algorithm [17].

In fitness landscape, the distance between a local optimal solution and a global optimal solution is defined as the minimal search steps from this local optimal solution to the global one by a local search algorithm. In practice, this distance is usually defined as Hamming distance [16]. The Hamming distance between solution $X$ and a global optimal solution $X^*$ is given by $d_H(X, X^*) = n - \left| X \cap X^* \right|$.

Figure 2 shows the fitness landscape of two classic NRP instances, nrp-1-0.5 and nrp-4-0.5. The details of the instances can be found in Section 5.1. For each sub-figure, the x-axis is the normalized Hamming distances (the Hamming distances divided by the scale of solutions) from the local optimal solutions to the global one and the y-axis is the normalized profits of the local optimal solutions (the profits divided by the profit of the global optimal solution).

feasible solutions and picks the best one out of these solutions; in (b) and (d), we obtain local optimal solutions by hill climbing algorithm [1]. Both algorithms run 1000 rounds and obtain 1000 local optimal solutions. For comparison, each algorithm in a sub-figure is respectively conducted with 1000 and 10000 iterations to find a local optimal solution. For example, an algorithm with 1000 iterations can provide local optimal solutions, each of which is the best one among 1000 solutions in its neighborhood.

As the fitness landscape shown in Figure 2, the distances between local optimal solutions and global optimal solutions are 0.30-0.60 times of the instance scale when using a randomized search algorithm for instance nrp-1-0.5 in (a); the distances are 0.45-0.60 times when using a hill climbing algorithm in (b). And for instance nrp-4-0.5, the distances are 0.42-0.52 times in (c) and 0.48-0.53 times in (d). This result indicates that there is a large overlap between local optimal solutions and global optimal solutions. In addition, for each sub-figure, the local optimal solutions with 10000 iterations tend to provide shorter distances than those with 1000 iterations. This result shows that a relatively strong local search algorithm may improve the local optimal solutions both on profits and on the similarity to the global optimal solutions.

## 4.2 Approximate Backbone and ABMA

The fitness landscape analysis in Section 4.1 shows that there is an overlap between local and global optimal solutions. Thus, we can approximate the backbone with the intersection of local optimal solutions. Given a set of local optimal solutions $F_L = \{X_1^L, X_2^L, \cdots, X_i^L\}$, the approximate backbone $a\_bone(F_L)$ is defined as the intersection of the local optimal solutions in $F_L$, i.e., $a\_bone(F_L) = X_1^L \cap X_2^L \cap \ldots \cap X_i^L$. Based on the approximate backbone, we design the ABMA algorithm. All the local optimal solutions in ABMA are obtained by a specified local search operator $H$.

Algorithm 1 shows the details of ABMA. The kernel operation of ABMA includes two phases: reduction and refinement. Every phase consists of multiple levels. A level in a multilevel algorithm is one step for reducing the instance scales or refining the solutions [20]. In the reduction phase, the algorithm first obtains the approximate backbone by the local search operator, then reduces the scale of the original instance to generate a new instance by fixing the approximate backbone, and solves the new instance at last. In the refinement phase, the algorithm combines the approximate backbone and the solution of the new instance together so as to construct a solution of the original instance. ABMA iteratively calls reductions and refinements. The number of iterations depends on the scale of the instance after reduction. In order to achieve high quality solutions, we also employ the multi-restart strategy in ABMA.

The advantage of ABMA is mainly attributed to its multiple reductions for instances. ABMA can dramatically reduce the search space of NRP to achieve high quality solutions by fixing the approximate backbone. Given an NRP instance with $n$ customers, if fixing the approximate backbone with scale $n'$, the upper limit for the scale of search space will be reduced from $2^n$ to $2^{n-n'}$. For example, when $n = 100$ and $n' = 30$, the upper limit will nearly decrease from $10^{30}$ to $10^{21}$. Since one reduction

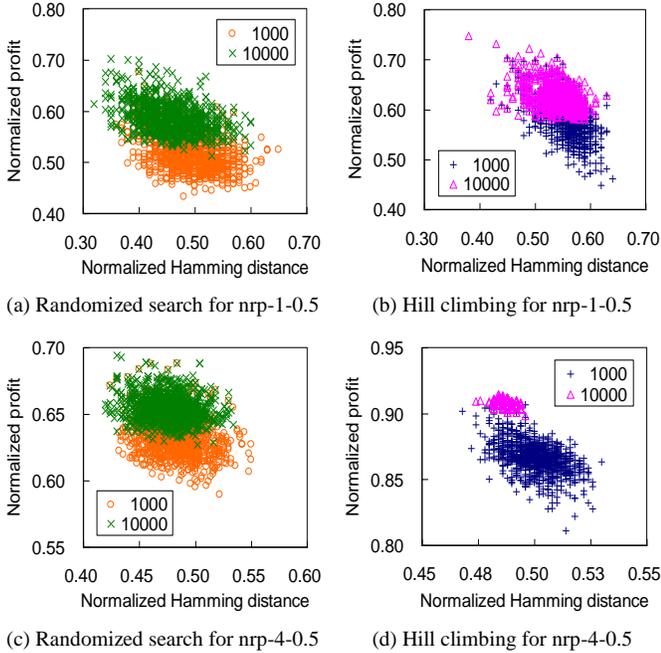

(a) Randomized search for nrp-1-0.5    (b) Hill climbing for nrp-1-0.5

(c) Randomized search for nrp-4-0.5    (d) Hill climbing for nrp-4-0.5

**Figure 2. Landscape of two NRP instances with two algorithms**

Among four sub-figures of Figure 2, we present the fitness landscape of instance nrp-1-0.5 in (a) and (b); and we present the fitness landscape of instance nrp-4-0.5 in (c) and (d). In (a) and (c), the local optimal solutions are obtained by the randomized search algorithm, which randomly generates a certain number of cannot reduce the large instance (e.g., $n = 500$) to a very small one, ABMA employs the multilevel strategy.

---

**Algorithm 1**: ABMA

**Input:**   instance $NRP(S,R,W)$, local search operator $H$,
            number of randomized restarting $\tau$

**Output:** solution $X*$

---

$\omega* = 0,\, d = 0$
**while** ($d < \tau$) **do**   //restart
(1)  $NRP(S_1, R_1, W) = NRP(S,R,W)$, $k = 1$ //initialize
(2) **while** ($|S_k|$ is large) **do**  //reduce the instances
   (2.1) Obtain the set of local optimal solutions $F_k$ by $H$
        for $NRP(S_k, R_k, W)$
   (2.2) Generate the approximate backbone $a\_bone(F_k)$
   (2.3) Reduce the original instance to $NRP(S_{k+1}, R_{k+1}, W)$, where
       $R_{k+1} \setminus (\bigcup_{(i,1) \in a\_bone(F_k)} \hat{R}_i)$,
       $S_{k+1} = S_k \setminus \{i | (i,1) \in a\_bone(F_k) \text{ or } (i,0) \in a\_bone(F_k)\}$
   (2.4) $k = k+1$
(3) Obtain local optimal solution $X$ by $H$ for new instance
   $NRP(S_k, R_k, W)$   //solve the small instance
(4) **while** ($k > 1$) **do**   //refine the solutions
   (4.1) $X = X \bigcup a\_bone(F_{k-1})$
   (4.2) $k = k-1$
(5) **if**($\omega* < \omega(X)$) **then** $X* = X$,  $\omega* = \omega(X)$ //update solutions
(6)  $d = d+1$

---

Figure 3 shows the reduction and refinement phases in ABMA for an NRP instance with 5 customers and 8 requirements. For this instance, the algorithm employs two-level reductions and refinements. In the first level reduction (Figure 3(a)), there are 5 customers and 8 requirements in the original instance. The local search operator obtains a set of 3 local optimal solutions $F_1 = \{X_1^1, X_2^1, X_3^1\}$. Thus, the first level approximate backbone is $a\_bone(F_1) = \{(2,1),(3,0)\}$. By fixing the approximate backbone $a\_bone(F_1)$, a new instance with 3 customers and 5 requirements is generated after reduction. For the second level reduction (Figure 3(b)), a set of 3 local optimal solutions $F_2 = \{X_1^2, X_2^2, X_3^2\}$ is obtained. Thus, the second level approximate backbone is $a\_bone(F_2) = \{(5,1)\}$. By fixing $a\_bone(F_2)$, a new instance with 2 customers and 2 requirements is generated (Figure 3(c)). For the local search operator, this instance is small enough to solve and the solution is $X = \{(1,1),(4,0)\}$. At last, under the inverted sequence of reduction, the algorithm combines the solution $X$ and the two approximate backbones together to construct a solution ($X = \{(1,1),(2,1),(3,0),(4,0),(5,1)\}$) for the original instance (Figure 3(d)).

## 4.3 Greedy Climbing Search

In ABMA, we employ a local search operator to generate local optimal solutions for achieving the approximate backbone and solving the small instance. Although any local search operator can be used in ABMA, a good one can improve the performance of the whole algorithm. In this section, we propose a greedy climbing search (GCS) operator for ABMA.

In the literature, the best local search algorithm for NRP is LMSA, a simulated annealing algorithm. As an extension to the stochastic hill climbing strategy, this algorithm controls the probability of accepting solutions by a temperature parameter. LMSA can work well on small scale instances of NRP. However, it may take too much time for solving large scale instances, due to the large search space [1][2]. Therefore, GCS operator is proposed to replace LMSA as the operator in ABMA. In contrast to LMSA, GCS is also an extension to the stochastic hill climbing strategy. GCS tends to choose the customers with high profits so as to obtain the good solution quickly from those randomly generated solutions.

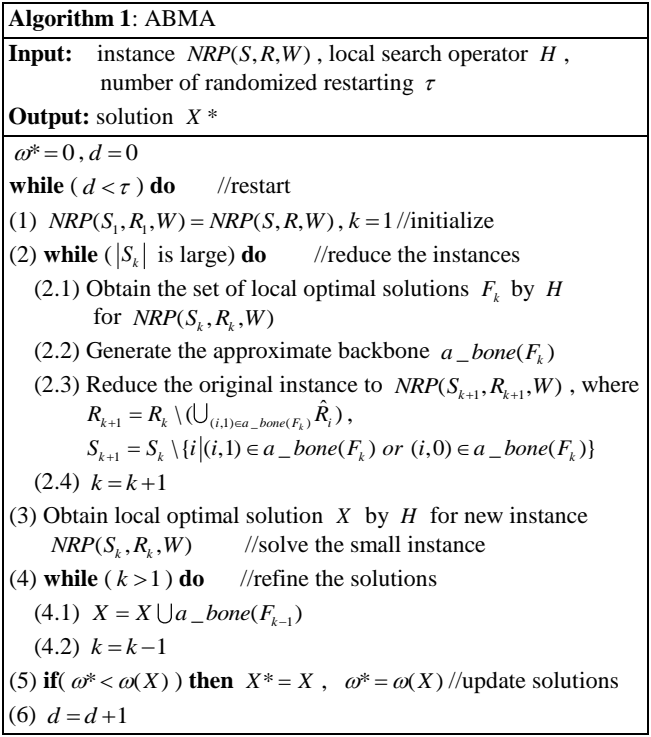

(a) The 1st level reduction

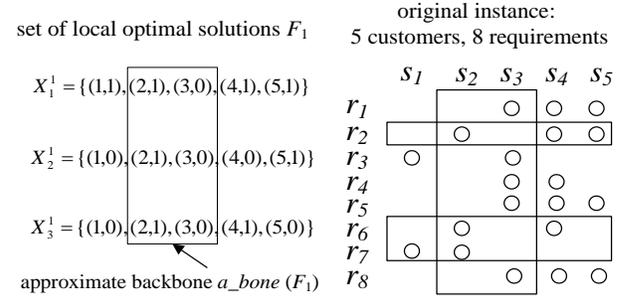

(b) The 2nd level reduction

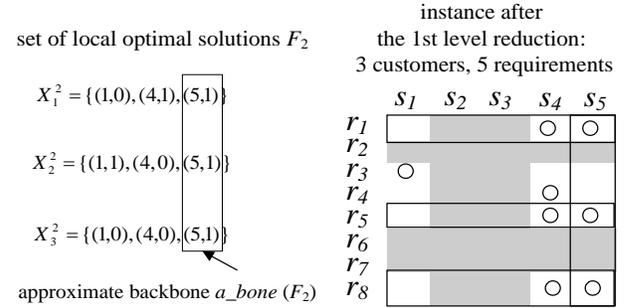

(c) Solving the small instance

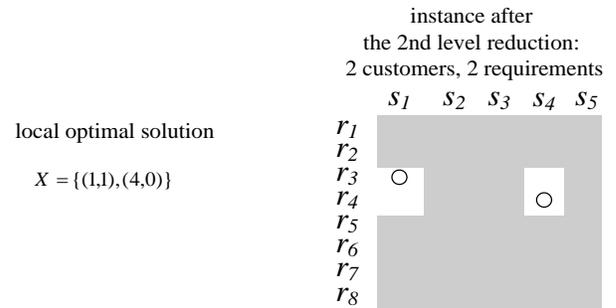

(d) Two levels of refinement

**Figure 3. Illustration of ABMA on an instance with 5**

**customers and 8 requirements**

Algorithm 2 shows the details of GCS, which consists of a series of iterations. In every iteration, if the solution is a feasible one, GCS adds a non-selected customer with maximum profit to the solution to enlarge the profit. Otherwise, GCS removes a selected customer randomly to construct a feasible solution. Since the time complexity for computing the cost of a solution is $O(nm^2)$ [1], the total time complexity of GCS is $O(\gamma nm^2)$, where $\gamma$ is the number of iterations. Therefore, ABMA with GCS operator is still a polynomial time algorithm.

---

**Algorithm 2: GCS**

**Input:**  instance $NRP(S, R, W)$, budget bound $B$,
number of iterations $\gamma$

**Output:** solution $X *$

Randomly generated solution $X$, $X* = X$, $i = 1$

**while** ( $i < \gamma$ ) **do**  //hill climbing and greedy strategy

(1) **if** ( $cost(X) < B$ ) **then** //increase profit for a feasible solution

  (1.1) Add a non-selected customer with the maximum profit,
  $X = (X \setminus \{(j, 0)\}) \bigcup \{(j, 1)\}$

  **else**  //change a non-feasible solution into a feasible one

  (1.2) Remove a selected customer randomly,
  $X = (X \setminus \{(j, 1)\}) \bigcup \{(j, 0)\}$

(2) **if** ( $\omega(X*) < \omega(X)$ ) **then**  $X* = X$      //update the solution

(3) $i = i + 1$

---

## 5. EXPERIMENTAL RESULTS

For approximate algorithms, it is the common way to evaluate the performance of algorithms by experimental methods. In the experiments of this paper, the algorithms are implemented with $C++$, compiled under $g++$, and run on a PC with *Intel Core 2.53 GHz* processor and *Fedora 6.0* OS (*Linux kernel 2.6*).

### 5.1 NRP Instance Generation

NRP is a classic problem arising from software requirement engineering. Since the requirement information is usually the privacy data of software companies, no open large NRP instances can be found in the literature. In this paper, we follow the classic literature of NRP experiments [1] to generate large NRP instance under certain constraints. These instances include 5 groups and every group includes 3 instances. In every group, instances have distinct budget bounds, i.e., the cost ratio (0.3, 0.5, and 0.7, respectively) multiplied by the sum of all costs.

Table 2 shows the details of constraints and all the values are non-negative integers. Taken the group of nrp-1 for example, all the requirements are classified into 3 levels separated by the symbol "/". A requirement in the 3rd level may depend on some requirements in the 2nd level. Similarly, a requirement in the 2nd level may depend on those in the 1st level. An instance name is formed by the group name and cost ratio. For example, nrp-1-0.3 is an instance in the group of nrp-1 and the budget bound is 0.3 multiplied by the sum of all costs. The details of instance nrp-1-0.3 are as follows. There are 3 levels of requirements, 20, 40, and 80 requirements in each level. The costs of requirements in the 1st level vary between 1~5, those in the 2nd level vary between 2~8, and those in the 3rd level vary between 5~10. A requirement in the 1st level may rely on at most 8 requirements in the 2nd level. Similarly, a requirement in the 2nd level may rely on 2 requirements in the 3rd level. There are 100 customers, with which 1~5 requirements are requested. In addition, every customer can provide a profit between 1~50.

### 5.2 Experimental Results and Analysis

To evaluate the performance of ABMA, we compare the experimental results of LMSA, GCS, and ABMA (with GCS operator) on the NRP instances in Table 3. There are 4 columns in this table. The first column is the detail of instances and the sub-columns are instance name, cost ratio, and budget bound, respectively. The other three columns are the experimental results of those 3 algorithms with 2 or 4 sub-columns. The sub-columns "profit" and "time" are the objective function values and time for computing in seconds. The sub-column "ratio" shows the ratio deviated from this algorithm to LMSA in percents. In more details, the ratio of profit is the percents improved on LMSA and the ratio of time is the percents increased on LMSA (the negative ratio is the percent reduced).

Since approximate algorithms need input parameters to control the process of algorithms, we set the parameters as follows. In LMSA (according to [1]), we set round of restarting to $10^2$, number of iterations to $10^6$ in each round at most, temperature to 0.01~0.3, and the temperature parameter to $10^{-8}$. In GCS, we set the round of restarting to $10^5$ and number of iterations to $10^3$ in each round at most. In ABMA, we set the round of restarting to $10^2$, number of iterations to $10^3$ in each round at most, and number of local optimal solutions to 10 for each approximate backbone in each reduction phase. Moreover, the reduction stops when the scale of the instance after reduction is less than 30% of the original one. We run all the algorithms for 10 times and record the averages of profits (accurate to integers) and time (accurate to 2 decimal places).

Table 3 demonstrates the experimental results of algorithms on NRP instances. It can be observed that LMSA achieves good performance on small instances and ABMA can work better on relatively large instances than the other two algorithms. The reason is that LMSA is a kind of simulated annealing algorithm, which can provide a good diversity in the search space. Especially on the instance nrp-1 with 100 customers and the instance nrp-2-0.3 with 500 customers, LMSA can obtain much better solutions than the other algorithms. Moreover, nrp-2-0.3 is also an easy instance for LMSA because the cost ratio decides the complexity of instances in the same group. Obviously, the cost ratio 0.3 will result in a smaller search space than the cost ratio 0.5. For other instances, ABMA can obtain better solutions than LMSA. This result can be attributed to the approximate backbone based on the similarity of the global and local optimal solutions. Under the guideline of the approximate backbone, ABMA reduces the scale of instances iteratively. Thus, ABMA tends to search the good local optimal solutions which are similar to global ones.

Each of GCS and LMSA cannot beat the other one in the comparison of running time while the time of both these algorithms depends on the number of iterations. ABMA takes less time than LMSA owing to the quick convergence in the search space by GCS operator. In addition, on the instances nrp-2-0.5 and nrp-4-0.3, ABMA takes much more time than LMSA. This result is caused by the number of iterations for obtaining the approximate backbone. When the instances are complex and the approximate backbones are hard to achieve, ABMA will run for numerous iterations. It will result in the sharp increase of running time.

**Table 2. Generation rules of NRP instances**

|  | nrp-1 | nrp-2 | nrp-3 | nrp-4 | nrp-5 |
|---|---|---|---|---|---|
| # of requirements per level | 20/40/80 | 20/40/80/160/320 | 250/500/750 | 250/500/750/1000/750 | 500/500/500 |
| cost of requirement | 1~5/2~8/5~10 | 1~5/2~7/3~9/4~10/5~15 | 1~5/2~8/5~10 | 1~5/2~7/3~9/4~10/5~15 | 1~3/2/3~5 |
| # of dependent requirements | 8/2/0 | 8/6/4/2/0 | 8/2/0 | 8/6/4/2/0 | 4/4/0 |
| # of customers | 100 | 500 | 500 | 750 | 1000 |
| # of requests of customer | 1~5 | 1~5 | 1~5 | 1~5 | 1 |
| profit of customer | 1~50 | 1~50 | 1~50 | 1~50 | 1~50 |

**Table 3. Performance: LMSA, GCS, and ABMA**

| instance | | | LMSA | | GCS | | | | ABMA | | | |
|---|---|---|---|---|---|---|---|---|---|---|---|---|
| name | ratio | bound | profit | time(s) | profit | ratio% | time(s) | ratio% | profit | ratio% | time(s) | ratio% |
| nrp-1-0.3 | 0.3 | 252 | <u>958</u> | 162.77 | 742 | -22.55 | 148.14 | -8.99 | 814 | -15.03 | 27.16 | -83.32 |
| nrp-1-0.5 | 0.5 | 421 | <u>1501</u> | 203.91 | 1240 | -17.39 | 178.77 | -12.33 | 1485 | -1.07 | 23.80 | -88.33 |
| nrp-1-0.7 | 0.7 | 589 | <u>2121</u> | 230.61 | 1874 | -11.65 | 222.33 | -3.59 | 2049 | -3.39 | 11.77 | -94.90 |
| nrp-2-0.3 | 0.3 | 1517 | <u>3122</u> | 688.30 | 2263 | -27.51 | 691.19 | 0.42 | 2336 | -25.18 | 217.39 | -68.42 |
| nrp-2-0.5 | 0.5 | 2529 | 5781 | 978.23 | 4432 | -23.34 | 882.88 | -9.75 | <u>6117</u> | 5.81 | 2319.78 | 137.14 |
| nrp-2-0.7 | 0.7 | 3540 | 7622 | 1164.53 | 7642 | 0.26 | 1095.30 | -5.94 | <u>7742</u> | 1.57 | 114.92 | -90.13 |
| nrp-3-0.3 | 0.3 | 2613 | 4820 | 1320.06 | 4886 | 1.37 | 1374.09 | 4.09 | <u>4926</u> | 2.20 | 737.05 | -44.17 |
| nrp-3-0.5 | 0.5 | 4355 | 8213 | 1704.30 | 9075 | 10.50 | 1828.16 | 7.27 | <u>9890</u> | 20.42 | 3051.45 | 79.04 |
| nrp-3-0.7 | 0.7 | 6096 | 12032 | 1788.89 | 12455 | 3.52 | 1941.08 | 8.51 | <u>12437</u> | 3.37 | 257.64 | -85.60 |
| nrp-4-0.3 | 0.3 | 6684 | 6059 | 2548.45 | 6685 | 10.33 | 2680.81 | 5.19 | <u>7342</u> | 21.18 | 11641.50 | 356.81 |
| nrp-4-0.5 | 0.5 | 11141 | 11688 | 3305.41 | 12721 | 8.84 | 3459.66 | 4.67 | <u>13040</u> | 11.57 | 2964.31 | -10.32 |
| nrp-4-0.7 | 0.7 | 15597 | 17701 | 3549.41 | 17954 | 1.43 | 3816.08 | 7.51 | <u>17970</u> | 1.52 | 1059.64 | -70.15 |
| nrp-5-0.3 | 0.3 | 1186 | <u>11710</u> | 2008.84 | 8277 | -29.32 | 1843.31 | -8.24 | 8255 | -29.50 | 1285.05 | -36.03 |
| nrp-5-0.5 | 0.5 | 1976 | 16495 | 2420.56 | 16145 | -2.12 | 2445.09 | 1.01 | <u>16673</u> | 1.08 | 953.89 | -60.59 |
| nrp-5-0.7 | 0.7 | 2766 | 24440 | 2430.76 | 24262 | -0.73 | 2789.83 | 14.77 | <u>24550</u> | 0.45 | 344.73 | -85.82 |

As mentioned above, the scales of search spaces depend on the cost ratio of instances in the same group. The results of LMSA and GCS show that the running time of these two algorithms increase along with the cost ratio growth. On the contrary, there is no such feature on the results of ABMA. Since it is affected by the instance characteristic, the running time of ABMA varies against the complexity of instances.

# 6. CONCLUSIONS AND FUTURE WORK

As an important problem in requirement engineering, NRP tries to balance the profits of customers and the costs for development. In this paper, we analyze the computational complexity for obtaining the backbone of NRP. We show that there exists no polynomial time algorithm to obtain the backbone of NRP instance under the assumption that $P \neq NP$. After the analysis of the relationship between local optimal solutions and global optimal solutions, we design ABMA to solve NRP by reducing large instances into smaller ones. Experimental results demonstrate that ABMA achieves good performance on the large NRP instances.

For NRP, our future work will focus on the estimation of the scale of the approximation backbone and some other approaches for obtaining the approximate backbone. First, the fitness landscape analysis shows that the scale of backbones relies on the instances and local search algorithms. However, the scale of approximate backbones is obtained by experiments. A further work is to explore the estimation for the scale of approximate backbones by theoretical analysis. Second, the backbone of NRP instances is constructed based on the intersection of solutions. Since the backbone can express some characteristics of problems, it is valuable to design better models to extract the backbone for algorithm design. We expect to obtain approximate backbones with a probability based model to add more customers to the backbone. This model may improve the diversity of solutions for NRP.

The idea in solving NRP can also be applied to other problems arising from the real-world applications in software engineering. We expect that the backbone based algorithms can be helpful to some other problems. Moreover, to date, there is no open instance library for NRP. We also want to collect several problem instances as an open library in our future work.

## 7. ACKNOWLEDGMENTS

We thank Yuanyuan Zhang at Department of Computer Science, King's College London for helpful suggestions. We thank our anonymous reviewers for valuable comments and corrections.

Our work is partially supported by the Natural Science Foundation of China under Grant No. 60805024, the National Research Foundation for the Doctoral Program of Higher Education of China under Grant No. 20070141020.